\documentstyle[amssymb,twocolumn,aps]{revtex}
\begin{document}
\title{Enhanced fluctuations of the tunneling density of states near bottoms of
Landau bands measured by a local spectrometer.}
\author{J.P. Holder$^1$, A.K. Savchenko$^1$, Vladimir I. Fal'ko$^2$, B. Jouault$^3$,
G. Faini$^3$, F. Laruelle$^3$, E. Bedel$^4$}
\address{$^{1}$ Department of Physics, University of Exeter, Stocker Road, Exeter,\\
EX4 4QL, UK\\
$^{2}$ Department of Physics, Lancaster University, Lancaster, LA1 4YB, UK\\
$^{3}$ L2M-CNRS, 196 Avenue H.Ravera, B.P. 107, F-92225 Bagneux, Cedex,\\
France\\
$^4$ LAAS-CNRS, Toulouse, Cedex, 31077, France}
\maketitle

\begin{abstract}
We have found that the local density of states fluctuations (LDOSF) in a
disordered metal, detected using an impurity in the barrier as a
spectrometer, undergo enhanced (with respect to SdH and dHvA effects)
oscillations in strong magnetic fields, $\omega _c\tau \geq 1$. We attribute
this to the dominant role of the states near bottoms of Landau bands which
give the major contribution to the LDOSF and are most strongly affected by
disorder. We also demonstrate that in intermediate fields the LDOSF increase
with $B$ in accordance with the results obtained in the diffusion
approximation.
\end{abstract}


Resonant tunneling through individual impurities has been identified and
studied in\ vertical \cite{Su,Dellow,Geim} and lateral \cite
{Fowler,McEuenn,Savch} mesoscopic structures. When an impurity level in a
potential barrier passes through the Fermi level in the emitter, it
manifests itself as a step in the current-voltage (IV) characteristic , with
the magnitude determined by the impurity coupling to the reservoirs and the
onset smeared due to the coupling or the thermal distribution of carriers in
the contact. Upon increasing bias, the current onset is followed by a
plateau where temperature-independent and magnetic-field-sensitive
reproducible features have been observed in several experiments on
small-area vertical structures \cite{Geim,Schmidt,Schmidt2}. The latter were
attributed to the\ fluctuations in the local density of single-particle
states \cite{Lerner}in a disordered emitter. It has then been suggested \cite
{Schmidt,Falko}that the impurity carrying the current (spectrometer) can act
as a probe of the local density of states fluctuations (LDOSF) in the bulk
of metallic contacts. When shifted with a bias, the spectrometer detects a
'fingerprint' of the LDOSF\ as a function of energy.

In the present paper, we study the evolution with magnetic field of the
LDOSF in a 3D disordered metal, a heavily doped semiconductor, and discuss
the results from the point of view of the fluctuation and correlation
properties of single-particle wave functions in disordered media. We have
measured the fingerprint of the LDOSF, $\delta \nu (\varepsilon )$, in the
differential conductance $G(V)=\frac{dI}{dV}(V)$ in a broad range of
magnetic fields, $B$, and analyzed its variance, $\left\langle \delta
G^2\right\rangle $, and correlation parameters. In intermediate fields, $%
\omega _c\tau \sim 1$, we have detected an increase of the fluctuation
magnitude, in agreement with the theoretically predicted behavior \cite
{Falko}: $\langle (\delta G)^2\rangle _B/\langle (\delta G)^2\rangle
_{B=0}\approx 1+(\omega _c\tau )^2$. At higher fields, $\omega _c\tau \geq 1$%
, we have observed large $1/B-$periodic oscillations in $\left\langle \delta
G^2\right\rangle $. We conclude that the LDOSF in strong fields are
dominated by the states near the bottoms of the Landau bands which have a
distinguished role relative to the rest of the spectrum.

The investigated structure consists of a 50\AA\ GaAs well imbedded between
two 81\AA\ Al$_{0.33}$Ga$_{0.67}$As barriers. Each Si doped GaAs contact
consists of three layers: 4800\AA\ with nominal doping 10$^{18}$cm$^{-3}$ is
followed by 4800\AA\ with 2$\times $10$^{17}$cm$^{-3}$ and the latter is
separated from the barrier by undoped spacer of 300\AA\ and 200\AA , for top
and bottom contact respectively. The lateral area of the nominally undoped
quantum well is reduced to a 700\AA\ diameter disk using the ion bombardment
technique \cite{Faini}. This decreases the number of active impurities in
the barrier, thus avoiding overlapping spectra of the LDOSF produced by
individual spectrometers. A schematic band diagram of the resonant tunneling
device with an impurity level $S$ in the quantum well is shown in Fig. 1,
inset. By testing several samples, we have selected one with a distinct
impurity level, which is also well separated from the states of the quantum
well which lie about 10 meV above. This energy range determines the interval
where the LDOS in the contact can be studied.

At zero bias, the spectrometer level $S$ is above the Fermi level $\mu $ of
the emitter with 3D metallic conduction{\bf .} The alignment of $S$ and $\mu 
$ with increasing bias is registered as a step in IV. In the differential
conductance $G(V)$ shown in Fig. 1, this current threshold corresponds to a
peak at 0.05 V. At low temperatures, its height is $G_\Gamma \simeq \frac{%
4e^2}h\frac{\Gamma _{\max }\Gamma _{\min }}{\Gamma _{\max }^2+\Gamma _{\min
}^2}\simeq $ $\frac{4e^2}h\frac{\Gamma _{\min }}{\Gamma _{\max }}$ and its
width is related to the energetic width of the spectrometer $\Gamma \simeq
\Gamma _{\max }$ $\simeq $ $120$ $\mu $eV determined by the tunneling
coupling between the impurity and the contacts \cite{ResTunn}. The values of 
$\Gamma _{\min ,\max }$ depend on the transparencies of the two barriers, so
that $\Gamma _{\max }$ corresponds to the lower (collector) barrier and $%
\Gamma _{\min }$ to the higher (emitter) barrier, $\Gamma _{\min }\sim
5\times 10^{-3}\Gamma _{\max }$ as estimated from the value of $G_\Gamma $.
The relation between bias $V$ and the energy scale of the spectrometer is
established by the coefficient $\beta =\frac{dE}{d(eV)}=0.24$, found for the
selected structure from the analysis of the temperature smearing of the
threshold peak.

Above the threshold, the current is determined by the emitter barrier
transparency $\Gamma _{\min }$ and the emitter density of states $\nu $ at
the energy $E_S$ below the Fermi level. As the barrier height does not
change significantly over a small $V$-range, the current becomes a measure
of the LDOS in the emitter: $I(V)\propto \nu (E_S)$. Fluctuations with
energy of the LDOS give rise to the reproducible, temperature independent
fine structure in IV. It is seen on top of a smooth decrease in the current
reflecting the averaged 3D density of states.

In some samples, we have detected Zeeman splitting of the spectrometer level
in magnetic fields parallel and perpendicular to the current. In such cases
it is seen that the upper spin level generates a replicated image of the
LDOSF which is shifted along the $V$-scale with respect to the image
produced by the lower level. We use this observation as a proof that two
levels of the same spectrometers probe the same pattern in the LDOSF. The
shift of the images corresponds to the difference between the spin-splitting
of an impurity level in the quantum well and that of free electrons in the
bulk. To avoid the overlap, a sample has been chosen with no significant
splitting of the images in magnetic field.

Fig. 2 shows the dependence $G(V)$ measured in magnetic fields $0<B<10.5$T
applied parallel to the current and changed with a step of 20 mT.
Fluctuations $\delta G(V)$ have a correlation voltage of $\Delta V_c\approx
0.5$ mV, which is comparable to the width $\Gamma /e\beta $ of the threshold
conductance peak. With increasing magnetic field up to $B\sim $ 4 T the
fingerprint in $G(V,B)$ changes randomly, with a correlation field $\Delta
B_c\thicksim $ 0.05T{\bf . }At high fields, the fluctuations transform into
a more regular pattern where individual features, assigned to Landau bands,
tend to move with increasing field towards the threshold peak, similar to
the observation by Schmidt {\it et al} \cite{Schmidt2}.

To interpret fluctuations $\delta G(V)$ as an image of LDOSF, we employ a
picture based on the properties of single-electron wave functions, $\psi _i(%
{\bf r})$ in a disordered metal \cite{FalkoEfetovRev,Mirlin,Blanter}. We
want to stress that the LDOSF measured by the resonant tunneling
spectroscopy reflect not only the randomness in the structure of chaotic
wave functions obeying the Porter-Thomas statistics \cite
{Brody,FalkoEfetovRev}, $\left\langle |\psi ({\bf r})|^{2n}\right\rangle
=f(n)L^{-nd}$ ($L$ and $d$ are the sample size and dimensionality, $f(n)=n!$
in weak fields and $\left\langle ...\right\rangle $ means the averaging over
disorder or energy), but also the existence of local correlations between
energetically close eigenstates \cite{Mirlin,Blanter}. In a phase-coherent
3D metal, the local density of states at a point ${\bf r}$ detected by
spectrometer with width $\Gamma $ can be considered as a sum of local
densities, $|\psi _i({\bf r})|^2$, of all eigenstates within energy interval 
$\Gamma $: 
\begin{equation}
I(V)\propto \nu (E)\sim \Gamma ^{-1}\sum_{|E_S-E_i|<\Gamma }|\psi _i({\bf r}%
)|^2.  \label{eq1}
\end{equation}
The sum in Eq. (\ref{eq1}) includes a large number of eigenstates, $N\left(
\Gamma ,L\right) \sim \nu _0\Gamma L^d$, each of which typically
contributing as little as $|\psi ({\bf r})|^2\sim L^{-d}$, with the mean
value of the LDOS, $\nu _0$, independent of the sample size $L$. As far as
fluctuations $\delta \nu $ are concerned, one might naively expect that
these fluctuations should vanish upon enlarging the sample, since for the
sum of $N\left( \Gamma ,L\right) $ independently fluctuating values $\delta
|\psi ({\bf r})|^2$ $\sim L^{-d}$ the variance $\left\langle \delta \nu
^2\right\rangle $ can be estimated as $N(\Gamma ,L)\left\langle \left(
\delta _L|\psi ({\bf r})|^2\right) ^2\right\rangle $, which is equivalent to 
$\Gamma ^{-2}\nu _0\Gamma L^dL^{-2d}\sim \nu _0\Gamma ^{-1}L^{-d}\rightarrow
0$ when $L\rightarrow \infty $. However, the correlations between wave
functions at close energies make the LDOSF in a large sample finite and
independent of its size. This statement can be explained using Thouless's
scaling picture of quantum diffusion \cite{Thouless}. We construct the
electron states in a large sample by representing them as linear
combinations of wave functions defined in its smaller parts, one of which
contains the observation point ${\bf r}$, and by gradually combining the
smaller parts up to the actual size $L$ of the sample. For an intermediate
length scale $\xi $ of the constituent part containing the observation
point, its states (which we call 'mother' states of generation $\xi $ ) are
spaced by $\Delta (\xi )\sim \left( \nu _0\xi ^d\right) ^{-1}$. Diffusive
spreading of these states into a larger part, when it is combined of several
blocks, leads to their random mixing with the states from the neighboring $%
\xi $-size blocks within the Thouless energy $\gamma \sim hD/\xi ^2$ \cite
{Thouless} , $D$ is the classical diffusion coefficient.

Since at each stage only a finite basis is involved in the construction of
the new states, some correlations exist between the new eigenstates,
although at small $\xi $ the spread of 'mother' states $\gamma $ is larger
than $\Gamma $ and these correlations are small. However, the Thouless
energy $\gamma $ decreases with increasing $\xi $ , and once $\xi $ exceeds
length $L_\Gamma =\sqrt{\hbar D/\Gamma }$ , the information carried by a set
of 'mother' states from a generation $\xi >L_\Gamma $ will not leave the
energy interval covered by the spectrometer. As a result, the sum of the
densities $|\psi _i({\bf r})|^2$ in Eq. (\ref{eq1}) will only depend on the
situation at the length scale $L_\Gamma $ and not vary with further refining
of the spectrum. Thus, $L_\Gamma $ and $N(\Gamma ,L_\Gamma )$ represent the
largest length scale and number of states for which correlations between
individual eigenfunctions could be neglected and the above naive estimate of 
$\left\langle \delta \nu ^2\right\rangle $ from independent fluctuators
used. Then, for the random difference between two values of $\nu $ in the
neighboring $\Gamma $-intervals one should take $\left\langle \delta \nu
^2\right\rangle \sim N(\Gamma ,L_\Gamma )\left\langle \left( \delta
_{L_\Gamma }|\psi ({\bf r})|^2\right) ^2\right\rangle \sim \nu _0\Gamma
^{-1}L_\Gamma ^{-d}$.

From this, we can now estimate the fluctuations in the current plateau
regime of the differential conductance which is a measure of the derivative,
with respect to energy, of the LDOS in Eq. (\ref{eq1}). We normalise the
variance $\left\langle \delta G^2\right\rangle $ by the height of the
threshold conductance peak which depends on the average LDOS and the
spectrometer width $V_\Gamma =\Gamma /\beta e$, so that $G_\Gamma \propto
N(\Gamma ,L_\Gamma )\left\langle |\psi ({\bf r})|^2\right\rangle /V_\Gamma $
. Because $\left\langle \delta G^2\right\rangle $ can be taken from the
above uncorrelated difference in $\nu $ as $\left\langle \delta
G^2\right\rangle \propto \left\langle \delta \nu ^2\right\rangle /V_\Gamma
^2 $, we arrive at 
\begin{equation}
\langle \delta G^2\rangle /G_\Gamma ^2\approx N(\Gamma ,L_\Gamma
)^{-1}\approx \left( \Gamma /hD\right) ^{(d-2)/2}/(\nu hD).  \label{eq3}
\end{equation}
We also estimate the correlation voltage of fluctuations as $V_\Gamma $ and
the correlation magnetic field as $\Delta B_c\simeq \Phi _0/L_\Gamma ^2$,
where $\Phi _0$ is the flux quantum.

In a 3D system with an anisotropic diffusion tensor $(D_x,D_y,D_z)$, Eq. (%
\ref{eq3}) transforms into $\langle \delta G^2\rangle /G_\Gamma ^2\propto
\left( D_xD_yD_z\right) ^{-1/2}$. This relation also determines the
classical effect on the variance $\langle \delta G^2\rangle $ of a magnetic
field ${\bf B}=B{\bf l}_z$. Assuming that the cyclotron motion suppresses
transverse diffusion as $D_{x,y}=D/(1+(\omega _c\tau )^2)$ \cite
{Stone,Maslov,Yosefin}, gives \cite{Falko}:

\begin{equation}
\langle \delta G^{2}\rangle _{B}/\langle \delta G^{2}\rangle _{B=0}\approx
1+(\omega _{c}\tau )^{2},\;\omega _{c}\tau \lesssim 1.  \label{eq4}
\end{equation}

Fig. 3 represents the result of our statistical analysis of conductance
fluctuations in small magnetic fields. \ The amplitude of fluctuations is
found from an individual $G(V)$ curve at a fixed $B$ in Fig. 2 as $%
\left\langle \delta G^2\right\rangle =\left\| (G(B,V)-\left\| G(B,V)\right\|
)^2\right\| $ ( $\left\| ...\right\| $ stands for the averaging over range $%
\Delta V$ $\approx $ 6 mV after the threshold peak ). To decrease the
scatter, a further averaging over a $B$-range of 0.25T has been performed.
The result is compared to that in Eq. (\ref{eq4}). The increase in $\langle
\delta G^2\rangle _B$ agrees with the expected quadratic dependence. From
Fig. 3, we find the momentum relaxation time $\tau \approx 0.9\times 10^{-13}
$s and use it to estimate the mobility, $\mu $ = 0.22 m$^2$/Vs, in the
emitter. The obtained values agree with those expected for the emitter with
the same nominal doping \cite{Poole}, and justify our use of the diffusion
approximation since $\tau \Gamma /\hbar \sim 10^{-2}$. We also use these
values to estimate the zero-field diffusion coefficient, $D\approx 40$ cm$^2$%
/Vs and $\Delta B_c\simeq \Gamma /eD\thicksim 0.03$T which is close to the
experimental value.

The value of $\tau $ confirms that the crossover from weak to strong fields, 
$\omega _c\tau \sim 1$, takes place at $B\approx $ 4T where the Landau band
(LB) formation is seen in Fig. 2. In the $\omega _c\tau \geq 1$ regime, the
field dependence of the variance $\left\langle \delta G^2\right\rangle $ has
a strong oscillatory character similar to de Haas-van Alphen (dHvA) effect,
with a sequence of peaks periodic in $1/B$ , Fig. 4a. However, the
oscillations in $\left\langle \delta G^2\right\rangle $ are much more
pronounced than the oscillations in the threshold peak $G_\Gamma $
reflecting the modulation of the average density of states at the Fermi
level in the emitter caused by depopulation of LB's, Fig.4b. Also, the
observed oscillations look significantly enhanced when compared to the
Shubnikov-de Haas (SdH) oscillations of conductance in a lateral GaAs
MESFET\ structure with the same nominal doping as the emitter, Fig. 4a,
inset. The positions of the peaks in the variance $\left\langle \delta
G^2\right\rangle (B)$ appear to be different from those in $G_\Gamma (B)$
and correspond to the filling of the LB's for the electron concentration of
approximately $3\times 10^{17}$cm$^{-3}$, which is larger than that near the
barrier and suggests that electrons in the emitter further from the barrier
could contribute to their origin.

These enhanced dHvA-type oscillations in the fluctuation amplitude suggest
that the above estimation of $\langle \delta G^2\rangle $ using statistical
properties of typical wave functions should be modified. This can be done by
considering a special role of the states with anomalously large fluctuations
of a local density, by analogy with \cite{FalkoEfetov} where these states
were 'prelocalized' states. In the case of a smooth random potential with
suppressed inter-LB scattering, these anomalous states are the states near
the bottoms of LB's.When in strong fields Abrikosov's dimerization of
electron motion \cite{Abrikosov} takes place, the contribution to the LDOSF
from the bottom of the highest filled LB becomes distinguished from typical
LDOSF and dominates in the magnitude of the variance $\langle \delta
G^2\rangle $. For energies $E_S$ close to the bottom of the $n$-th LB, $E_n$ 
$=(n+1/2)\hbar \omega _c$, not only transverse but also the longitudinal
diffusion coefficient related to the highest LB, $D_z^{(n)}\sim u_z^2\tau
\propto \left[ E_S-E_n\right] $, is suppressed due to the decrease in the
kinetic energy of the quasi-one-dimensional electron motion along magnetic
field. When the characteristic length scale $L_\Gamma ^z=\sqrt{\hbar
D_z^{(n)}/\Gamma }$ becomes smaller than the inter-LB scattering length, the
states from the upper LB start providing a contribution $\langle \delta
^{(n)}G^2\rangle $ to the LDOSF that is enhanced compared to the typical
variance $\langle \delta ^{(typ)}G^2\rangle $:

\begin{equation}
\langle \delta ^{(n)}G^2\rangle /\langle \delta ^{(typ)}G^2\rangle \approx
\left( \nu ^{(n)}/\nu \right) \left( D_0/D_z^{(n)}\right) ^{1/2}  \label{eq5}
\end{equation}
The structure of Eq. (\ref{eq5}) explains the enhancement of oscillations in
the fluctuation amplitude in Fig.4 relative to oscillations of $G_\Gamma (B)$
and SdH oscillations. The latter are the measure of the ratio of the LDOS in
the highest ($n-$th) LB and the total LDOS, i.e. they are represented by the
first factor in Eq. (\ref{eq5}). The unusual factor $\left(
D_0/D_z^{(n)}\right) ^{1/2}$, which is responsible for the enhancement of
the oscillations of the fluctuation amplitude, is a specific feature of the
LDOSF effect.

We thank I. Lerner, R.J. Haug, T. Schmidt and A.D. Stone for discussions and
EPSRC for support.

\begin{figure}[tbp]
\caption{Differential conductance as a function of bias with the threshold
peak and the 'fingerprint' of the LDOS below the Fermi level in the emitter.
Inset: Band diagram of the resonant tunneling structure with a spectrometer.}
\label{Fig. 1}
\end{figure}

\begin{figure}[tbp]
\caption{Conductance fluctuations $G(V,B)$ normalised to the threshold peak.
Curves for different $B$ are offset upwards and multiplied by an increasing
factor to compensate for the decrease of the threshold peak with field.}
\label{Fig. 2}
\end{figure}

\begin{figure}[tbp]
\caption{a) Diagram of the emitter volume where the tunneling LDOS is
formed, at $B=0$ and $B>0$. $L_\Gamma $ is the diffusion length
corresponding to electron life-time $\hbar /\Gamma $ at the impurity level.
b) Increase of the conductance fluctuations in intermediate fields due to
the suppression of transverse diffusion.}
\label{Fig. 3}
\end{figure}

\begin{figure}[tbp]
\caption{a) Oscillations of the conductance variance in strong fields.
Inset: SdH oscillations in the bulk conductivity. b) For comparison, magneto
oscillations of the threshold conductance peak.}
\label{Fig. 4}
\end{figure}

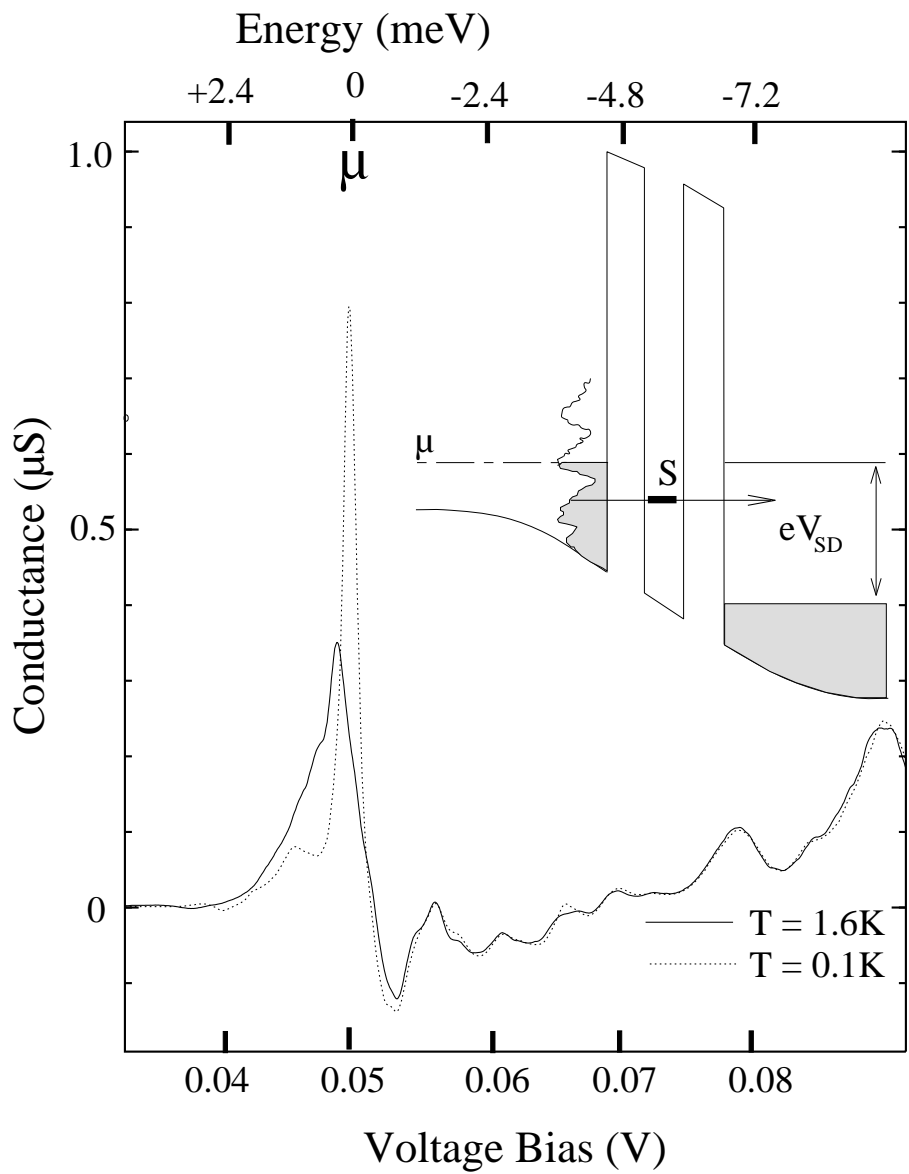
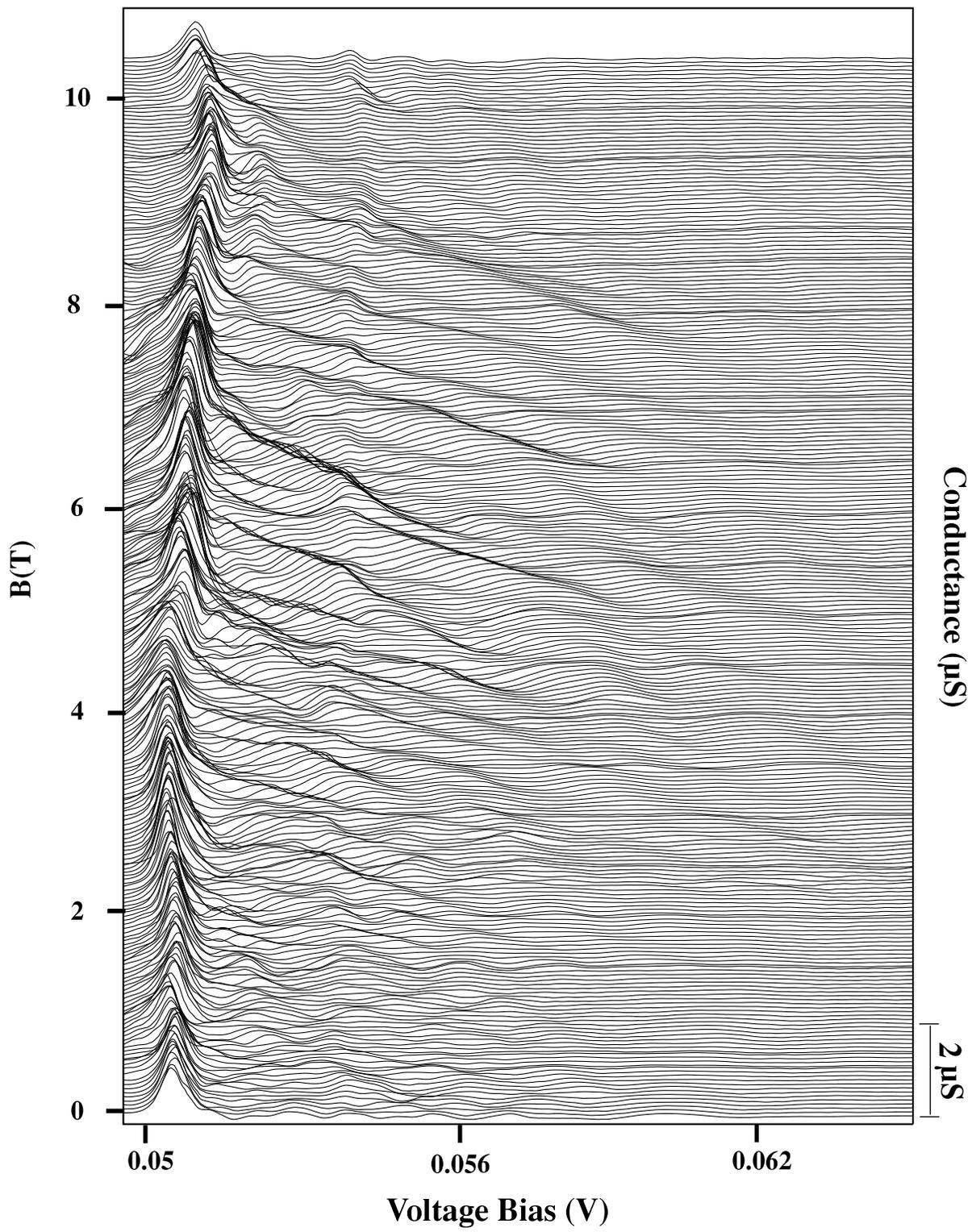
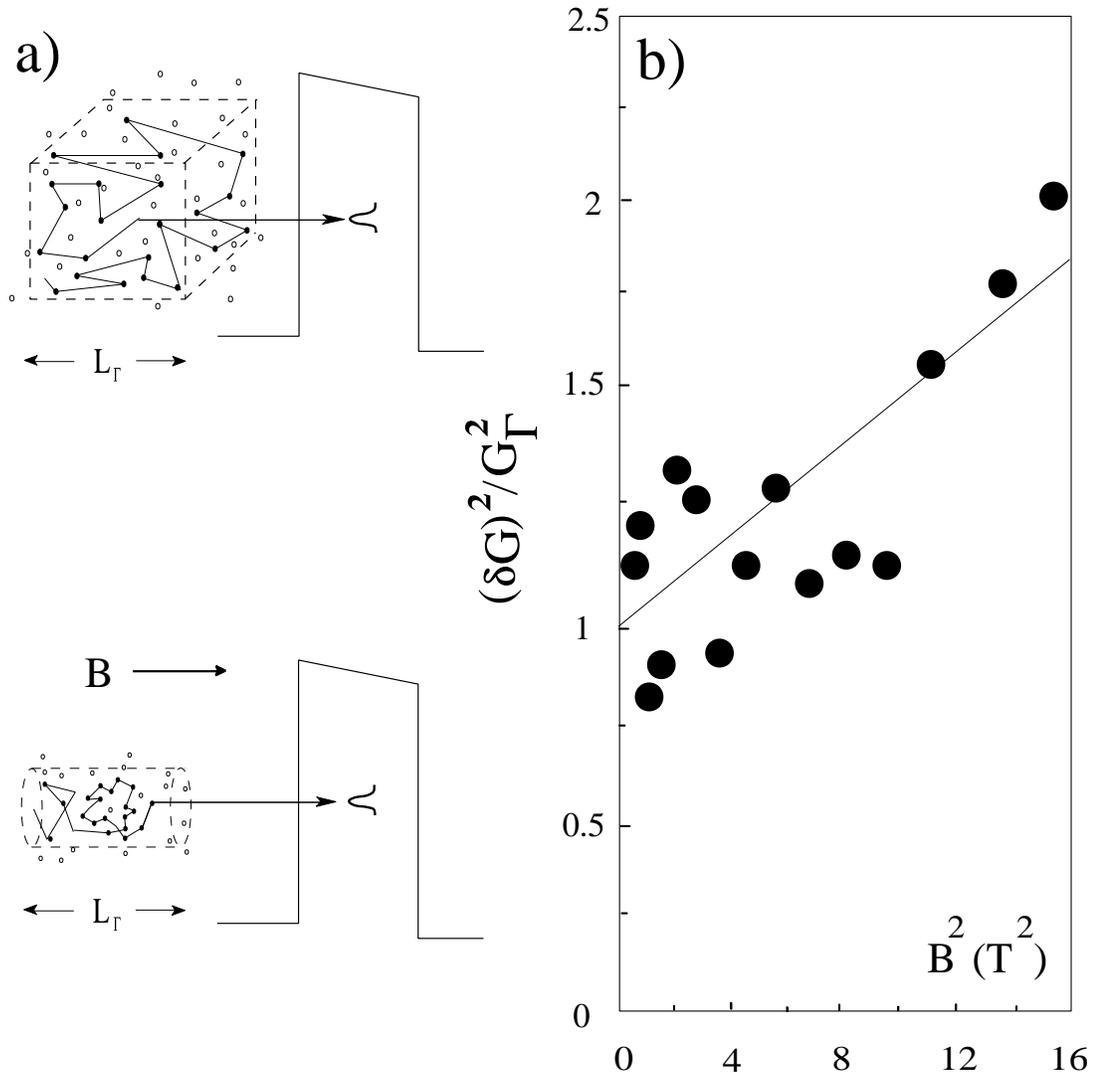
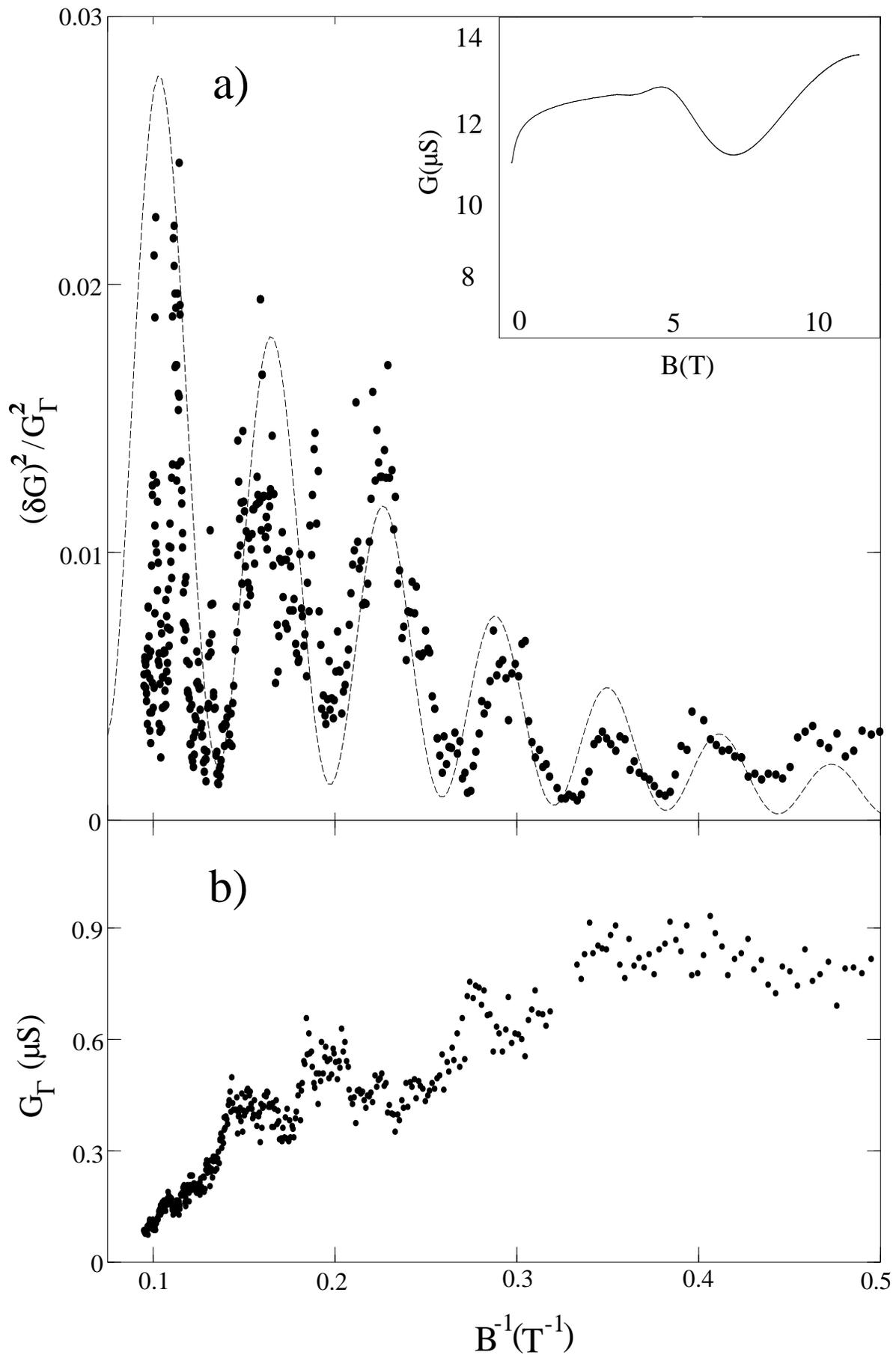
\end{document}